\documentclass[12pt]{article}
\usepackage{amsfonts,epsfig}
\usepackage{graphicx}
\usepackage{amsmath,amsfonts}
\usepackage{amssymb,graphicx}
\def\pa{\partial}
\def\s{\sigma}

\def\d{\delta}
\def\h{\hat}

\def\dda{\ddot{a}}

\def\e{\varepsilon}
\def\a{\alpha}
\def\g{\gamma}
\def\m{\mu}
\def\n{\nu}
\def\be{\begin{equation}} 
\def\ee{\end{equation}}  
\def\ba{\begin{eqnarray}}
\def\ea{\end{eqnarray}}  

\def\h{\hat}
\def\b{\bar}
\def\t{\tilde}
\begin{document}  
\begin{flushright}
\end{flushright}
\vspace{10pt}
\begin{center}
  {\LARGE \bf Cosmological constraints on parameters of one-brane models
with extra dimension} \\
\vspace{20pt}
Mikhail Z.Iofa\\
\vspace{15pt}

\textit{Skobeltsyn Institute of Nuclear Physics, 
Moscow State University, Moscow,119992 , Russia}\\

    \end{center}
    \vspace{5pt}
\begin{abstract}
We study some aspects of cosmologies in 5D models with one infinite 
extra dimension. Matter is confined to the brane, gravity extends to the 
bulk.
Models with positive and negative tension of the
brane are considered. Cosmological evolution 
of the 4D world is described by warped 
solutions of the generalized Friedmann equation.
Cosmological
solutions on the brane are obtained with the input of the present-time
observational cosmological parameters. We estimate the 
age of the Universe and abundance of ${}^4 
He$ produced in primordial nucleosynthesis in different models. Using 
these estimates we find constraints on
dimensionless combinations of the 5D gravitational scale, scale of the 
warp factor and coupling at the 4D curvature term in the action. 
\end{abstract}
\section{Introduction}
Cosmological models with extra dimensions provide an extension of the
standard cosmological model and naturally appear in
 microscopic theories unifying gravity with other interactions.
As models of the real world they must be tested as cosmological 
theories.
In this work we study some aspects of cosmologies in 5D models with one 
infinite extra dimension. Matter is confined to the brane, gravity 
extends to the bulk.
 
The action of the model is
\be
\label{i1}
I_5  =\int\limits_\Sigma \,\sqrt{-g^{(5)}} \left(\frac{R^{(5)}}{2\kappa^2} +
\Lambda\right) +\int\limits_{\pa {\Sigma}}
\,\sqrt{-g^{(4)}}\left(\frac{R^{(4)}}{2\kappa^2_1} -\h\s \right) -
\int\limits_{\pa
{\Sigma}} \sqrt{-g^{(4)}}L_m
,\ee
 where $\kappa^2 =8\pi/M^{3}$, $\kappa^2_1 =8\pi/ r_c M^3$.
$L_m$ is the Lagrangian of matter on the brane. 
$M$ is the gravitational scale of the 5D gravity, parameter $r_c$ 
defines
the strength of gravity in the 4D term.

Cosmological evolution on the brane is described by solutions of the 
non-standard Friedmann equation \cite{RS1, 
RS2,BDL1,BDL2,Shtanov1,Shtanov2, maart}.
 In the Gaussian normal frame the metric in the bulk is 
found as a warped 
solution of the Einstein equations, satisfying  Israel junction
conditions on the brane with matter.

We consider models with positive and negative tensions of the
brane.
The latter case is suggested by models with two branes, which allow for 
a
possibility to reconsider the hierarchy problem. In these models the 
visible brane should have negative tension \cite{rub}.

In this work we do not make a fit of observational cosmological data,
but
taking as the input the set of present-time cosmological parameters
(Hubble
parameter, fractions of total energy density of cold matter and
radiation
and deceleration parameter ) we look for constraints on 
dimensionless combinations of scales
 of the models which follow from the requirement that
the models
reproduce the age of the Universe and  abundance of ${}^4 He$ produced
in primordial nucleosynthesis.

There are no strong constraints on parameters of the models with extra
dimensions. Models with the fundamental scale varying in a very broad 
range of values were considered.

In the framework of the ADD models with large extra dimensions 
\cite{ADD,AADD} the
fundamental scale of theory was taken at the scale of the
standard model  $M_{SM}$
thus evading the hierarchy problem.

A model with the action \cite{DGKN2}
$$
S=M^3 \int\,d^4 x\int^R_0 \, dy \sqrt{-g^{(5)}} R^{(5)} +M^3 r_c
\int \, d^4 x \sqrt{-g^{(4)}}R^{(4)}
,$$
with compactified extra dimension
with the fundamental scale $M$ in the $TeV$ range, compactification 
radius
$R\geq 10^{16} m \,\, (R\sim 10^{32} GeV^{-1})$ and $R/r_c \sim 10^{-4}$ 
was argued to be compatible with measurements of the Newton law and 
cosmological data.

In a number of papers (for a list of refs. see \cite{lue})
cosmology was investigated in different variants of
the DGP model \cite{DGP}. In these models the fundamental scale was 
taken well below the standard model scale.

In the model proposed in  \cite{DGKN3}
with the action of the form (\ref{i1}) with infinite flat extra 
dimension
 the fundamental scale is renormalized by the standard model
interactions to the level $\sim 10^{-3} eV$.
The model was argued to be consistent with collider experiments, 
cosmology and gravity measurements.

In papers \cite{def,DDG,DLRZA}  for asymptotically flat
metrics  and in \cite{lu.st} for dS metrics (in the static coordinate
system)
it was found that with the fundamental scale $M$ in the range $\sim 10\, 
- \, 10^2
MeV$ and $r_c\sim H^{-1}_0$ the models are consistent with observational 
data.
In \cite{DDG} this was verified with the input of parameters of
the standard cosmological model, in \cite{DLRZA} was performed more 
precise comparison of the model with the SN1 and CMB data with the 
independent fit
of parameters of the model (however in \cite{alsah} difficulties in
confronting the DGP model with SN data were reported).

In IIB string theory the RS scenario can be modeled as a 
stack of branes at the orbifold fixed
point with the warping scale  $\m\sim M_{st}/(4\pi gN)^{1/4}$, where
$M_{st}$
and $g$ are the string scale and the string coupling \cite{kraus}.
The scale $\m$ can be made arbitrarily small
for large enough number $N$ of stacked branes.
Identifying the scale $\m$ with the inverse
radius $R$ of extra dimension,
 a constraint $\m \geq 10^{-3} eV$ was
obtained from estimation of corrections to the Newton law in the
RSII model \cite{chung.ever.dav}.

Although collider physics predictions  were mostly discussed within the
ADD -like
models with the number of extra dimensions $n\geq 2$ and  with the
fundamental scale in the $TeV$ range  (for example,
\cite{hew,neerv,cheung}), it was noted
that many results remain qualitatively relevant
also for models with warped compactifications with
the fundamental scale of the same order as in the ADD model.

In the present paper, in models with the action (\ref{i1}), instead of 
asymptotically flat solutions we 
consider warped metrics (in the Gaussian normal frame).
The fundamental scale $M$ and the scale $\m$ of the exponent in the warp
factor of the metric {\it a'priory} are practically
unconstrained. In this situation we assume that $\m$ is in the range
$10^{-12} GeV\, - \, 10^{3} GeV$. The lower limit  is suggested by
precision of
measurements of the Newton law,
the upper limit is of order of the Standard Model scale.

The non-standard Friedmann equation in the models with extra dimension
contains
terms quadratic in total energy density and the dark radiation term
\cite{BDL1,BDL2,Shtanov1,Shtanov2}.
To obtain the estimates of the age of the Universe and of abundance of
Helium produced in primordial nucleosynthesis it is sufficient to
consider "late" cosmology,
i.e. times at which the terms quadratic in energy
densities are small as compared to the term linear in energy
densities.

Comparing predictions of the models with experimental data, we obtain a
number of relations connecting dimensionless combinations of
parameters of the models. In particular, we find relations
$$
\frac{\m M^2_{pl}}{M^3}\simeq \frac{2(1+q_0 )}{3\Omega_m}(\m
r_c \pm 1),
$$
where $q_0$ and
$\Omega_m$ are present-time acceleration parameter and the fraction of
cold matter in the
total energy density and the sign corresponds to the sign of tension of 
the
brane.

The paper is organized as follows. After briefly reviewing in Sect.1
 the one-brane
model, in Sect.2 we discuss the model with positive tension of the 
brane.
Two cases are considered separately: those with and without 4D curvature
term present in the 5D action. We calculate the age of the Universe and
find constraints on the scales of the models which follow from the 
requirement that the lifetime of the Universe is within the 
observational bounds..

In Sect.4 the same program is carried out for models with negative
tension of the brane. In this case, in distinction to the models with
positive tension, for different signs of the combination $\m r_c -1$ the 
pictures are
different. The model with $1>\m r_c$ does not yield correct age of the 
Universe, the model with $\m r_c >1$ can 
reproduce the observational data.

In Sect.5 we consider the model without dark radiation term in Friedmann
equation.

In Sect.6 we calculate abundance of Helium produced in the BBN  in the
models with positive and negative tensions. More precise 
constraints on parameters of the models are obtained.

In Conclusions we summarize constraints on parameters obtained in
different models.

\section{One-brane model}

We start with a brief introduction of 5D models  with the action 
(\ref{i1}).
It is assumed that 
matter is confined to the spatially flat brane and  gravity is 5-dimensional. 
To have homogeneous cosmology on the brane the energy-momentum 
tensor of matter on the brane is taken 
in a phenomenological form
\be
\label{2}
 T^\m_\n =diag\{ -\h{\rho}, \h{p}, \h{p}, \h{p}\},
\ee
where $\h{\rho }(t)$ and $\h{p}(t)$ are the sums of energy densities and
pressures of cold matter and radiation.
For the following it is convenient to introduce
the normalized expressions for bulk cosmological constant, energy
density, pressure and
cosmological constant on the brane which all have the same
dimensionality $GeV$
\be  
\label{4}
\m =\sqrt{\frac{\kappa^2\Lambda}{6}},
\quad \rho =\frac{\kappa^2\h\rho}{6},    
\quad \s =\frac{\kappa^2\h\s}{6},\quad {p} =\frac{\kappa^2\h{p}}{6}.
\ee
The Einstein equations are solved in a class of metrics of the form
\be
\label{3}
 ds^2_5 = -n^2 (y,t)dt^2 + a^2 (y,t)\g_{ij}dx^i dx^j +dy^2 .
\ee
The brane is located at the fixed position $y=0$. Using the freedom in
parametrization of time, the component of the metric $n(y,t)$ is normalized
so that $n(0,t)=1$. 

We study cosmologies on the brane by solving the non-standard Friedmann
equation \cite{BDL1,BDL2,Shtanov2,maart}
 which follows from the system of the Einstein equations and junction
(Israel) conditions on the brane (in the following we assume the symmetry
$y\rightarrow -y$)
\ba  
\label{5}
&{}& \frac{a'(0,t)}{a(0,t)}= -\s -\sum\rho_i (t)
+\frac{r_c}{2}\frac{\dot{a}^2(0,t)}{a^2(0,t)},
\\\nonumber
&{}& \frac{n'(0,t)}{n(0,t)} =-\s +\sum (2\rho_i +3p_i ) +\frac{r_c}{2}
\left(-\frac{\dot{a}^2(0,t)}{a^2(0,t)}
+2\frac{\ddot{a}(0,t)}{a(0,t)}\right), 
\ea
 where $\rho_i , p_i \,\, i=r,m $ are densities and pressures of the
cold matter and radiation.
Equation for the scale factor $a(0,t)$ can be written in the form with the
second-order
derivatives of the scale factor \cite{BDL1},
\be  
\label{6}
\frac{\ddot{a}(0,t)}{a(0,t)} +\frac{\dot{a}^2 (0,t)}{a^2 (0,t)} = -2\mu^2 -
\left(\s + \sum\rho_{i} -\frac{r_c}{2}\frac{\dot{a}^2 (0,t)}{a^2 (0,t)}\right)  
\left(-2\s + \sum\rho_{i} +3 \sum p_{i} + r_c\frac{\ddot{a}(0,t)}{a(0,t)}
\right)
,\ee
 or, starting from the partially
integrated system of the Einstein equations, in the form with the
first-order derivatives \cite{BDL2,Shtanov1,Shtanov2}
\be
\label{8}
\frac{\dot{a}^2 (0,t)}{a^2 (0,t)}=-\m^2 +
\left(\s + \sum\rho_{i} -\frac{r_c}{2}\frac{\dot{a}^2 (0,t)}{a^2
(0,t)}\right)^2 + C\frac{a^4 (0, t_0 )}{a^4 (0, t)}
,\ee
or, equivalently,
\be
\label{8a}
(1+\s r_c )H^2 = \s^2-\m^2 +2\s (\rho_m +\rho_r ) + \left(\rho_m +\rho_r -\frac{r_c
H^2}{2}\right)^2 + C(z+1)^4
\ee
where 
$$H(t)=\dot{a} (0,t)/a (0,t)  
 ,\,\,\,z =a(0, t_0 )/a(0, t)-1
$$
and
\be
\label{11}
\rho_m (z) =\rho_{m0} (1+z)^3, \quad \rho_r (z) =\rho_{r0} r(z) (1+z)^4
.\ee 
The function $r(z)$ is a slow function of $z$ which counts, as the 
function
$g_* (T)$ in the expression for radiation energy
density as the function of temperature, $\hat{\rho}_r =\pi^2 g_* (T)T^4
/30$, the number
of relativistic degrees of freedom.

C is the integration constant of dimension 
$mass^2$,  the term 
$$C(z+1)^4\equiv\m\rho_w (z)
$$ 
is interpreted as dark radiation \cite{maeda,maart}. 
$$
q_0 =-\frac{\dda(t_0)}{a(t_0 )H_0^2}
$$ 
is the present-time deceleration parameter.
 Equation (\ref{6}) can be
obtained from (\ref{8}) by differentiation over $t$ and the use of
conservation equation for energy densities of cold matter and radiation. 

\section{Model with positive tension of the brane}
\subsection{Model without 4D curvature term}
In this model the term $R^{(4)}/2\kappa_1^2$ in the action (\ref{i1}) is absent.
Setting in (\ref{8}) and (\ref{6})  $t=t_0$
 we have
\footnote{
Substituting the explicit $z$-dependent expressions for energy
densities and pressures in Eq.(\ref{6}) and integrating it we obtain
$$
H^2 =  (\s^2 -\m^2 )
 +2\s \rho_{m0}(z+1)^{3}  
 + (z+1)^{6}(\rho_{m0}+\rho_{r0}(z+1))^2 +\t{C}(z+1)^4 ,
$$
where $\t{C}$ is the integration constant. Comparing this equation with
(\ref{8}), we obtain the relation between the constants $C$ and $\t{C}:
\,\,\, C=\t{C} -2 \s\rho_{r0}$.}
\be
\label{14}
H^2_0 =\s^2 -\mu^2 +2\s (\rho_{m0} +\rho_{r0}) +(\rho_{m0} +\rho_{r0})^2 +
{C}
.\ee
\be
\label{15}
(1-q_0 )H^2_0 =2(\s^2 -\mu^2 ) +\s \rho_{m0}
-(\rho_{m0}^2+3\rho_{r0}\rho_{m0} + \rho_{r0}^2 )
.\ee
We shall consider the period of "late cosmology", when the terms linear in
energy densities are dominant in Friedmann equation 
\be
\label{lc}
\s > \rho_m (z) +\rho_r (z).
\ee
From relations (\ref{14}) and (\ref{15}), 
neglecting the terms quadratic in energy
densities,  we obtain
\be
\label{16}
\s^2 -\m^2=\frac{H^2_0}{2p^2}\left(-1+ (1-q_0 )p^2\right) 
,\ee
and
\be
\label{17}
 {C}\simeq \frac{H^2_0}{2p^2}\left(-3+ (1+q_0 )p^2 -4\Omega_r /\Omega_m\right)
.\ee    
Here we introduced the dimensionless ratio
\be
\label{9}
 p^2=
\frac{H^2_0}{\s\rho_{m0}}
.\ee
Using (\ref{16}) and (\ref{17}) we present Eq.(\ref{8}) in a form
\ba
\label{18}
H^2\simeq (z+1)^4 \frac{H^2_0}{2p^2}\left[
(-1+(1-q_0 )p^2 )(z+1)^{-4}+4(z+1)^{-1}\right.\\\nonumber
\left.+
\frac{H^2_0}{p^2 \m^2}(z+1)^2 \left(1+\frac{\Omega_{r}}{\Omega_{m}}
r(z) (z+1)\right)^2 -3+(1+q_0 )p^2 +\frac{\Omega_{r}}{\Omega_{m}}(r(z) 
-1) \right].
\ea
Neglecting the terms quadratic in energy densities and setting $r(z)=1$, 
we obtain the Friedmann
equation as
\be
\label{17a}
H^2\simeq \frac{H^2_0}{2}\left[1-q_0 +(1+q_0 )(z+1)^4 +\frac{1}
{p^2}(-1+4(z+1)^3 -3(z+1)^4 )\right]
.\ee
 The  rhs of (\ref{17a}) is positive for all $z$ if
\footnote{Although above we neglected the terms quadratic in
matter(radiation) densities, it is verified that the bound remains
valid with inclusion of these terms.}  
\be
\label{bou}
 (1+q_0 )p^2 -3 >0.
\ee
From (\ref{16}) we obtain that
\be
\label{27}
\frac{\s}{\m}=1 +O\left(\m r_c\frac{H^2_0}{\m^2}\right)
.\ee 
Below we assume that $(\m r_c)H^2_0/\m^2\ll 1$. 
Because $\s/\m\simeq 1$, we can write approximately
 $$
p^2 \simeq\frac{H^2_0}{\m\rho_{m0}}.
$$
Introducing
$$
\h{\Omega}_{V}=\frac{\s^2 -\m^2}{H^2_0},\quad
\h{\Omega}_{m} =\frac{2\m\rho_{m 0}}{H^2_0}=
\frac{\m M^2_{pl}}{M^3}\Omega_m,
\quad
\h{\Omega}_{r} =\frac{2\m\rho_{r 0}}{H^2_0}=
\frac{\m M^2_{pl}}{M^3}\Omega_r,
\quad\h{\Omega}_{d} =\frac{C}{H^2_0}
,$$
where $\Omega_{m,r}=\h{\rho}_{m,r 0}/{\h{\rho}_c}$, and 
$\hat{\rho}_c = 3H_0^2 M^2_{pl}/8\pi$ is the critical density of the
Universe approximately equal to the total present-time energy density, 
 we can present Eq. (\ref{14}) without the quadratic terms in a form
$$
1=\h{\Omega}_{V}+ \h{\Omega}_{m} +\h{\Omega}_{r} +\h{\Omega}_{d}.
$$
If we  set $\m M^2_{pl}/M^3 =1$ and $C=0$, we obtain an equation 
similar to
that in the standard 4D cosmology with $\s^2 -\m^2$ interpreted as the
dark energy term. However, in the present case neither $C$, nor $\m$ and 
$\s$ are fixed {\it a'priori}. 

To have a qualitative picture 
let us consider the form of the Friedmann equation in different
regions of $z$.
 
(i) Cold matter-dominated region
$1<z+1<\left({\Omega_{m}}/{\Omega_{r}}\right)\sim 10^{4}$,
where  Eq.(\ref{18}) is approximated by (\ref{17a}).

(ii) Radiation-dominated region where all
$z$-dependent terms are small as compared to the radiation term 
$-3 +(1+q_0 )p^2 =O(\Omega_r )$
\ba
\label{rd}
 &{}&  \frac{H^2_0}{p^2
\mu^2}z^2\left(1+\frac{\Omega_{r}}{\Omega_{m}}z\right)^2 \ll (1+q_0 )p^2 -3
 \\\nonumber
&{}&4z^{-1}\ll (1+q_0 )p^2 -3
.\ea
Conditions (\ref{rd}) are fulfilled for
\be
\label{rd1}
\Omega_r^{-1}\ll z<\left(\frac{p^4 \Omega_m^2 \m^2}{H^2_0 \Omega_r}\right)^{1/4}
\simeq 10^{23}(\m/GeV)^{1/2}
.\ee
In this region Eq.(\ref{18}) approximately is
\be                             
\label{r2}                          
H^2 =\frac{\dot{z}^2}{z^2}\simeq z^4 \frac{H^2_0}{2p^2} ( (1+q_0 )p^2 -3)
.\ee         
Condition (\ref{lc}) defining the period of late cosmology can be written as
\be
\label{bou2}
z  < \left(\frac{ p^2\Omega_m \m^2}{H^2_0 \Omega_r}\right)^{1/4},
\ee   
which is approximately the same as (\ref{rd1}).

(iii)       Radiation-dominated high-energy  region
$z>10^{25}(-3 +(1+q_0)p^2)^{1/4} p^{1/4}$,
where the term quadratic in densities is dominant, and Friedmann 
equation takes the form  
\be                                      
\label{r3}                                   
\dot{z}^2 \simeq \rho_{r0}^2 r^2 (z)z^{10}
\ee                                         
In both regions (ii) and (iii) one can use an approximate equation
  \be              
\label{r4}           
\dot{z}^2 \simeq z^6\left[\frac{(1+q_0 )p^2 -3 }{2}
 \m\rho_{m0} +\rho_{r0}^2 r^2 (z) z^4\right]=
z^6\frac{H^2_0}{2p^2}\left[-3+(1+q_0 )p^2
+\frac{H^2_0}{p^2\m^2}\left(\frac{\Omega_r }{\Omega_m} r(z)
z^2\right)^2\right]
\ee     

\subsection{The age of the Universe}
Integrating  Eq.(\ref{17a}), we obtain
\be
\label{r7}
 H_0 (t_0 -t)
=\frac{p}{\sqrt{2}} \int\limits^1_{(z (t)+1)^{-2}}\frac{du}
{\left[(-1+(1-q_0 )p^2 )u^2 +4u^{1/2} -3+(1+q_0 )p^2\right]^{1/2}},
\ee
where the bound (\ref{bou}) is understood.
To estimate of the age of the Universe we set $t=0$.
 For  $|q_0 | > 0.5$ the results are presented in Fig.1
\footnote{
For estimates we take
$H_0 = 10^{-42}\,GeV^{-1},\, q_0= -0.57,\,
\Omega_m =0.24,\,\,\Omega_r =4.6\cdot
10^{-5}$.}.
\vspace*{2.5cm}
\begin{figure}[ht]
\begin{picture}(0,0)(0,210)
\put(20,75){\epsfig{file =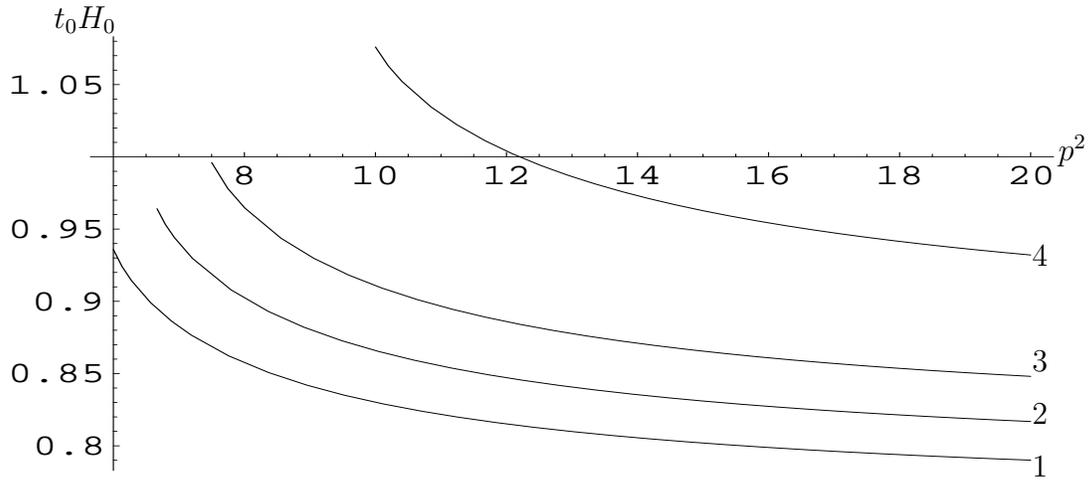 ,width =400pt, height =180pt}}
\put(40,250){$t_0 H_0$}
\put(420,200){$p^2$}   
\put(410,160){4}       
\put(410,120){3}       
\put(410,100){2}        
\put(410,80){1}        
\end{picture}          
\vspace*{5cm}          
\caption{$t_0 H_0$ as a function of $p^2$ for $|q_0| =
0.5,\,0.55,\,0.6,\,0.7$, curves 1-4.}
\end{figure}

From analysis of experimental data it follows
 that the normalized age of the Universe $t_0 H_0$ is close
to unity ( $t_0 H_0 =0.95 \pm 0.08$, 
for a review and refs. see \cite{steig}, recent processing of data
\cite{kom,rei,tegmark,seljak},
related discussion \cite{hsu}).
From Fig.1 it is seen that for $-q_0$ in the range  
$ 0.5\div 0.6 $ 
\footnote{For example, \cite{steig} give $-q_0 =0.58\pm 0.01$, \cite{harun}
  $-q_0 = 0.56\pm 0.04$. }
and for $p^2 \simeq 3/(q_0 +1)$, 
one obtains $t_0 H_0\simeq 0.94\div 0.96$. 
 For larger $|q_0|$ to obtain 
$t_0 H_0 \sim 1$ one must take larger larger $p^2$.
 
Because $\rho_{m0}\simeq \rho_c \Omega_m \simeq H^2_0 
M^2_{pl}\Omega_m/2M^3$, we can write 
$$
p^2\simeq \frac{2M^3}{\m M^2_{pl}\Omega_m}
.$$ 
Relation $p^2\simeq 3/(1+q_0 )$ can be equivalently written  as
\be
\label{r8}
\frac{\m M^2_{pl}}{M^3}\simeq \frac{2(1+q_0 )}{3\Omega_m}.
\ee
The rhs of (\ref{r8}) is close to unity 
\footnote{
If $q_0 \simeq -1 +3\Omega_m /2 $ \cite{part,harun,steig}, one obtains
$3\Omega_m/2(1+q_0 )\simeq 1$.}.
This is a natural result, 
because  with $\m M^2_{pl}/M^3\simeq 1$ 
 the model is close to the standard cosmological model (cf.
\cite{sahni}). 
For
 $\m \sim 10^3 \, GeV$ and $\m \sim 10^{-12} \, GeV$ we obtain  5D scales
 $M\sim 10^{14}GeV$ and $M\sim 5\cdot 10^8\, GeV$.

\subsection{Model with 4D curvature term in the 5D
action}
In this subsection we discuss solutions of the Friedmann equation
in the model with curvature term on the brane included in the 5D action
(\ref{i1}). 

As in the model with $r_c =0$, in the present model
it is useful to define the period of “late cosmology”. In the most
straightforward way
"late cosmology" can be defined as a period in which in the Friedmann
Eq.(\ref{8a})
the terms linear in matter (radiation) energy densities are dominant
\be
\label{lc1}
\s\rho(z) > \rho^2 (z), \quad \s\rho(z) > r_c H^2 \rho(z), \quad \s\rho(z) > (r_c
H^2 )^2 .
\ee
Additionally we assume that parameter $r_c$ is constrained so that
\be
\label{lc3}
 (\m r_c) \frac{H^2_0}{\m^2}\ll 1
.\ee
In the most stringent case, for $\m\sim 10^{-12} GeV$, this condition is 
valid for $\m r_c < 10^{60}$.
From Eqs. (\ref{6}) and (\ref{8}) taken at present time, neglecting
quadratic terms, we have
\be
\label{n4a}
\s^2-\m^2 \simeq \frac{H^2_0}{2p^2}\left[(1+r_c\s  )(1-q_0 )p^2 -1 \right]
,\ee    
and     
\be     
\label{n8}  
C\simeq 
\frac{H^2_0}{2p^2}\left[-3-4\frac{\Omega_r}{\Omega_m} +
(1+r_c\s )(1+q_0 )p^2 \right],
\ee
Below from the estimates of the age of the Universe it will be shown 
that
$$
p^2 (1+\m r_c )(1+q_0 )-3 \ll 1
,$$
and thus $p^2\simeq 3/(1+\m r_c )(1+q_0 )$.
Using in (\ref{n4a}) this relation, we have
$$
\left(\frac{\s}{\m}\right)^2 -1 \simeq \frac{H^2_0 
}{6\m^2}\left[(3(1-q_0 )- (1+q_0 )) +\m r_c \left(3(1-q_0 
)\frac{\s}{\m}- (1+q_0 )\right)\right]
.$$
Taking into account the bound (\ref{lc3}), we obtain that
\be
\label{lc4}
\frac{\s}{\m} =1 +O\left(\m r_c\frac{ H^2_0}{\m^2}\right)
.\ee
In the following, in expressions containing the factor $H^2_0/\m^2$, we set
$\s =\m$.
 
Provided conditions (\ref{lc1}) and (\ref{lc3}) are satisfied,
Friedmann Eq. (\ref{8a}) in the period of late cosmology can be 
approximately written as
\be
\label{lca}
(1+r_c \m )H^2 \simeq \s^2 -\m^2 +2\m (\rho_m (z) +\rho_r (z) ) +\m\rho_w (z)
.\ee
The first condition
 (\ref{lc1}), $\s\simeq\m > \rho (z)$,  
is valid for 
\be
\label{lc8}
z^4 \ll \frac{\m^2 p^2 \Omega_m }{\Omega_r H^2_0},
\ee
which is the same as condition (\ref{rd1}).
The second and the third conditions (\ref{lc1}) are satisfied, if the first
condition is valid.  

Substituting in (\ref{lca}) expressions (\ref{n4a}) and (\ref{n8}), we 
obtain
\be   
\label{n9}
H^2\simeq \frac{H^2_0}{2}\left[1-q_0 +(1+q_0 )(z+1)^4 +\frac{1}
{p^2 (1+\m r_c )}(-1+4(z+1)^3 -3(z+1)^4 )\right]
.\ee
which is the same as (\ref{17a})
in the case without 4D curvature term up to the 
 substitution
$$
p^2 \rightarrow  p^2 (1+ \m r_c ).
$$
The rhs of (\ref{n9}) is positive if
\be    
\label{n10}
p^2 (1+ \m r_c  )(1+q_0 )>3
.\ee
Making in the expression for the age of the Universe obtained in the 
model  with $r_c >0$ redefinition
 $p^2 \rightarrow p^2 (\m r_c +1)$, we obtain the corresponding 
expression in the model with $r_c >0$.

The normalized age of the Universe as a function of $|q_0|$ in the
region $0.5 - 0.6$ and with $\b{p}^2 = 3/(1+q_0 )$ is shown in Fig 2.
\vspace*{2cm}
\begin{figure}[ht]
\begin{picture}(0,0)(0,210)
\vspace*{4cm}
\put(20,75){\epsfig{file = 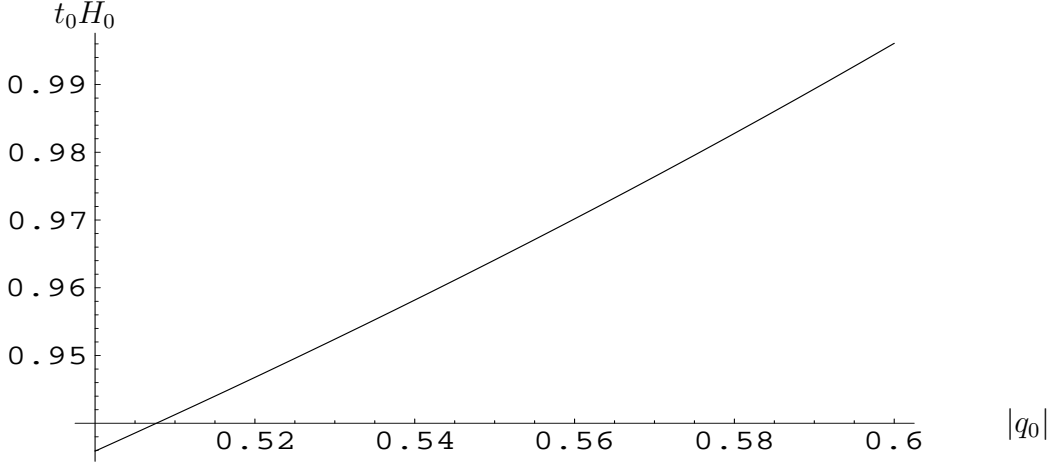 ,width =350pt, height =180pt}}
\put(40,250){$t_0 H_0$}
\put(400,95){$|q_0|$}  
\end{picture}
\vspace*{4.5cm}
\caption{$t_0 H_0$ as a function of $|q_0|$ with $p^2 =3/(1+q_0 )(1+\m r_c
)$ }
\end{figure}
\vspace*{-1.5cm}

\vspace*{2.5cm}
\begin{figure*}[h]
\begin{picture}(0,0)(0,210)
\put(20,75){\epsfig{file = 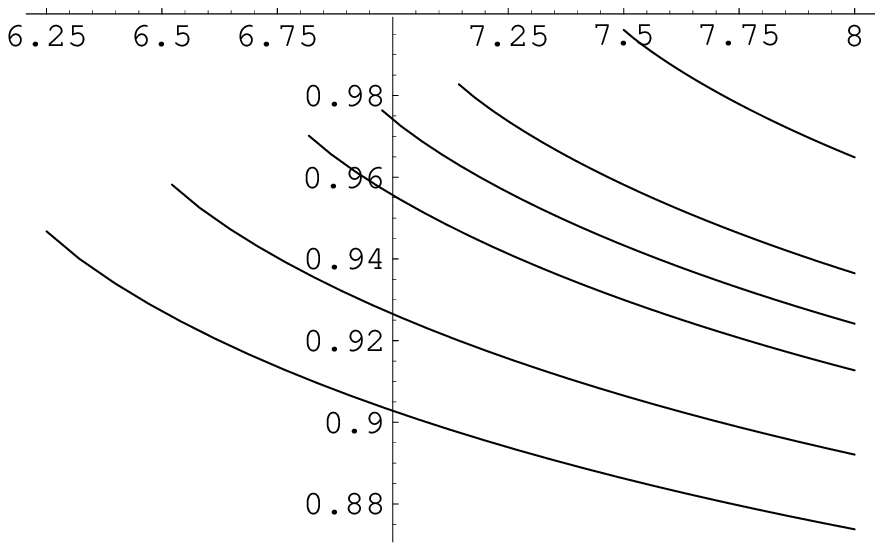
 ,width =350pt, height =180pt}}
\put(160,70){$t_0 H_0$}
\put(370,250){$p^2 (\m r_c +1)$}
\put(370,150){$\uparrow |q_0|$}
\end{picture}  
\vspace*{5.5cm}
\caption{$t_0 H_0$ as a function of $p^2 (\m r_c +1)$ for
$|q_0|=0.52,0.54,0.56,0.57,0.58,0.60$. The left end points
 of the curves are at $p^2 (\m r_c +1) =3/(1+q_0 )$. }
\end{figure*}

It is seen that the normalized age of the Universe $t_0 H_0$ is of order
unity, if $p^2 (\m r_c +1) \simeq3/(1+q_0 )$. 
Substituting $p^2 \simeq  2M^3/\m M^2_{pl}\Omega_m$,   we obtain
\be
\label{n11a}
\frac{\m M^2_{pl}}{M^3} \simeq \frac{2(1+q_0 )}{3\Omega_m}(\m
r_c +1),
\ee
where the combination $ 3\Omega_m/2(1+q_0 ) $ is of order unity.
\section{Model with negative tension of the brane}
\subsection{Model without 4D curvature term in the
5D action}
In the model with negative tension on the brane, using Eqs.
(\ref{14}) and (\ref{15}) with $\s =-|\s|$, we have
\ba
\label{p2}
(1-q_0 )H^2_0 =2(\s^2 -\mu^2 ) -|\s| \rho_{m0}
-(\rho_{m0}^2+3\rho_{r0}\rho_{m0} + \rho_{r0}^2 ),\\\l{p2a}
H^2_0 =\s^2 -\mu^2 -2|\s| (\rho_{m0} +\rho_{r0}) +(\rho_{m0} +\rho_{r0})^2 +
{C}
.\ea
Neglecting the terms quadratic in energy densities 
from these equations we express $\s^2 -\m^2$ and $C$
\ba
\label{p30}
\s^2 -\m^2 \simeq\frac{H^2_0}{2p^2}(1+(1-q_0)p^2 ),\\
\label{p31}
C\simeq\frac{H^2_0}{2p^2}\left[3+4\frac{\Omega_r}{\Omega_m} +
(1+q_0 )p^2 \right].
\ea   
Substituting (\ref{p30}) and (\ref{p31}), we obtain the Friedmann equation
as 
\ba   
\label{p40}
H^2\simeq (z+1)^4 \frac{H^2_0}{2p^2}\left[
(1+(1-q_0 )p^2 )(z+1)^{-4}-4(z+1)^{-1}\right.\\\nonumber
\left.+
\frac{H^2_0}{p^2 \m^2}(z+1)^2 \left(1+\frac{\Omega_{r}}{\Omega_{m}}
r(z) (z+1)\right)^2 +3+(1+q_0 )p^2 \right].
\ea
We neglected the term $(r(z) -1)\Omega_{r}/\Omega_{m}$ as compared with 
$3+(1+q_0 )p^2$.
In contrast to the model with positive tension of the brane, in the present
case we cannot make direct comparison of Eq. (\ref{p40}) 
with the corresponding relation in the
standard cosmological model. 
However, comparison of the age 
of the Universe and  production of Helium in BBN 
with the standard cosmological model is possible.

As in the model with positive tension, 
we distinguish the following regions of $z$.

(i) Cold matter-dominated region
$1<z+1<\left({\Omega_{m}}/{\Omega_{r}}\right)\sim 10^{4}$,
where  Eq.(\ref{p40}) can be approximated as
\be
\label{p5}
H^2 \simeq \frac{H^2_0}{2 p^2}\left[\left
({1 +(1-q_0 )p^2} \right) 
-4(z+1)^{3} + \left({3+ (1+q_0 )p^2}\right )(z+1)^4 \right]
\ee
For $q_0$ in the  physically interesting region $|q_0 | <1$
the rhs of (\ref{p5}) is a positive increasing function 
for all $z >0$, and there are no
constraints on $p^2$.

(ii) Radiation-dominated region, in which $z>{\Omega_{m}}/{\Omega_{r}}$ 
and 
$$
3+(1+q_0)p^2 > \frac{H^2_0\Omega_{r}^2}{p^2\m^2\Omega_{m}^2}z^4 
.$$
In this region Eq.(\ref{p40}) can be written as
\be                             
\label{p60}                          
H^2 \simeq z^4 H^2_0 \frac{3 +(1+q_0 )p^2 }{2p^2}.
\ee 
\newpage
        
(iii)       
Radiation-dominated high-energy  region
$z>10^{25} p^{1/2}$,
where the term quadratic in densities is dominant, and equation is
\be   
\label{p70}
\dot{z}^2 \simeq \rho_{r0}^2 r^2(z) z^{10}
.\ee                              
\vspace*{-1cm}
\subsection{The age of the Universe}
Integrating  Eq.(\ref{p5}), we obtain
\be
\label{r7a}
 H_0 (t_0 -t)
=\frac{p}{\sqrt{2}} \int\limits^1_{(z (t)+1)^{-2}}\frac{du}
{\left[(1+(1-q_0 )p^2 )u^2 -4u^{1/2} +3+(1+q_0 )p^2\right]^{1/2}}.
\ee
\vspace*{2.5cm}
\begin{figure*}[h]
\begin{picture}(0,0)(0,210)
\put(20,75){\epsfig{file =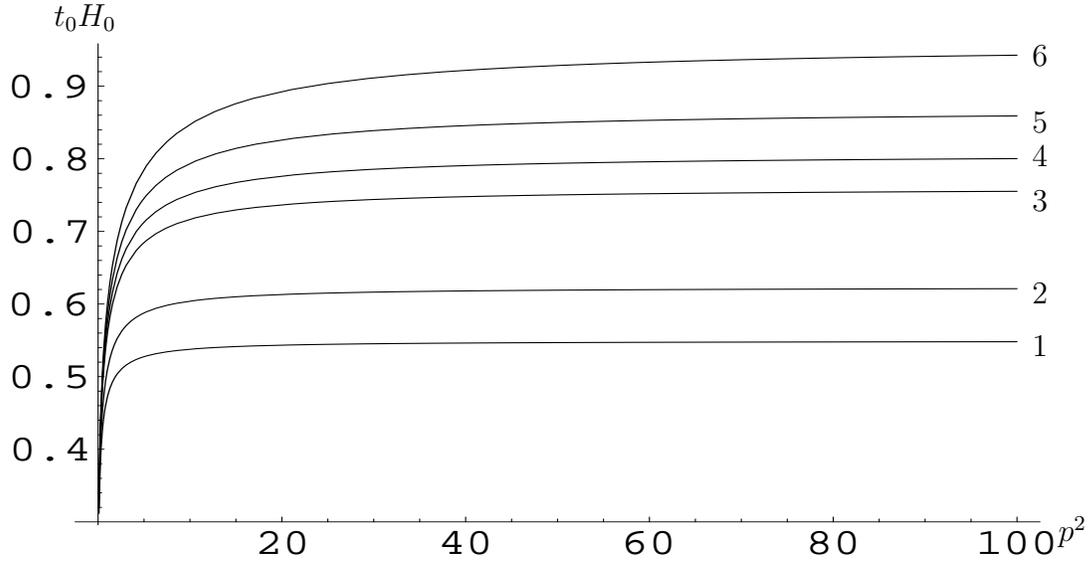 ,width =400pt, height =200pt}}
\put(40,280){$t_0 H_0$}
\put(420,85){$p^2$}
\put(410,265){6}
\put(410,240){5}
\put(410,227){4}
\put(410,210){3}
\put(410,175){2}
\put(410,155){1}
\end{picture}
\vspace*{5cm}
\caption{$t_0 H_0$ as a function of $p^2$ for $q_0 =
0.5,\,0,\,-0.5,\,-0.6,\,-0.7,\,-0.8$, curves 1-6.}
\end{figure*}
For an estimate  we have set $t=0$ and the lower limit in (\ref{r7a}) to
zero.
For $p^2 \simeq  2M^3/\m M^2_{pl}\Omega_m\ll 1$, or equivalently, for 
$\m M^2_{pl}/M^3\gg 1$,
  the normalized age of the Universe
\be   
\label{r6}
H_0 t_0 \simeq \frac{p}{\sqrt{3}}\left[\ln\frac{2\sqrt{3}}{p} -1 \right]
,\ee
is much below the currently accepted estimate $H_0 t_0\simeq 1$.

As $p$ increases, $t_0 H_0$ also increases. 
For $p^2\gg 1$ we find that $H_0 t_0$ is approximately independent of $p$.
For $|q_0 |\sim 0.5 - 0.6$ the age of the 
Universe is below the
observational bounds and smaller than in  $\Lambda$CDM  model with   
$\Omega_m \simeq 0.3$ and $\Omega_m
+\Omega_\Lambda =1$. To obtain $t_0 H_0\simeq 1$ one must take $|q_0|>0.9$
what is outside the allowed region of $q_0$ \cite{steig,part}.

\subsection{Model with 4D curvature term in the 5D
action}

Taking  Eqs. (\ref{14}) and (\ref{15}) at present time 
and  omitting the terms of higher order in energy densities, we have
\ba 
\label{n2b}
H^2_0 (1- r_c |\s| )(1-q_0) \simeq 2(\s^2 - \m^2 ) -|\s| {\rho}_{m0}, \\
\label{n3b}
H^2_0 (1- r_c |\s| ) \simeq
-\m^2 + \s^2  -2|\s| ({\rho}_{m0} +\rho_{r0} )  +C
.\ea 
From these relations we express $\s^2 -\m^2$ and $C$.
\ba
\label{n4b}
\s^2 -\m^2 \simeq\frac{H^2_0}{ 2p^2}\left(1+(1-q_0 )(1-\s r_c  )p^2\right),\\
C\simeq\frac{H^2_0}{2p^2}\left[3+4\frac{\Omega_r}{\Omega_m} +
(1-\s r_c  )(1+q_0 )p^2 \right]
.\ea
Conditions of late cosmology are the same as (\ref{lc1}) with the 
substitution $\s \rightarrow |\s|$. 
 Constraining $r_c$  so that $ (\m r_c) H^2_0/\m^2 \ll 1$, we show that with
this accuracy we can substitute $\m$ for $\s$.

If $1-\m r_c >0$, Friedmann equation in the period of late cosmology is
\be
\label{5a}
H^2\simeq \frac{H^2_0}{2}\left[(1-q_0 ) +(1+q_0 )(1+z)^4
+\frac{1}{{p}^2 (1-\m r_c )}\left(1-4(z+1)^3 +3(z+1)^4 \right)\right]
,\ee
which is the the same as (\ref{p5}) with substitution $p^2 \rightarrow
{p}^2 (1-\m r_c)$. As it was discussed in the previous subsection, this model does
not yield correct age of the Universe.
 
Let us consider the model with $\m r_c -1 >0$. 
In this case we obtain the Friedmann equation
in a form similar to Eq. (\ref{n9}) in the model with positive tension 
of the
brane
\be     
\label{n6b}
H^2\simeq \frac{H^2_0}{2}\left[1-q_0 +(1+q_0 )(z+1)^4 +\frac{1}
{p^2 (\m r_c -1)}\left(-1+4(z+1)^3 -3(z+1)^4 \right)\right]
.\ee
The rhs of (\ref{n6b}) is positive for
\be    
\label{n7b}
p^2 (\m r_c -1 )(1+q_0 )>3
.\ee
In this model, as in the model with positive tension of the brane, we
obtain the normalized  age of the Universe $t_0 H_0 \simeq 1$ if  
$p^2 (\m r_c -1  )
\simeq 3/(1+q_0 )$. This condition can be expressed as
\be
\label{n7c}
\frac{\m M^2_{pl}}{M^3} \simeq \frac{2(1+q_0 )}{3\Omega_m}(\m
r_c -1) 
.\ee 

\section{ Friedmann equation without the dark radiation term}
In the model with curvature term on the brane included in the 5D action,
the Friedmann equation
without the dark radiation term  $C/a^4 (0,t)$ takes the
form
\be   
\label{n11}
H^2 =\left(\rho +\s -\frac{r_c H^2}{2}\right)^2 -\m^2
.\ee 
From this equation written at the current-time we determine the tension of
the brane $\s$ as a function of $H^2_0$ and $\rho_0$, or equivalently, of
$p^2$. The deceleration parameter  will be discussed later.

As above, we  consider the period of late cosmology, when conditions
(\ref{lc1}) are valid.  Eq.(\ref{n11}) can be approximately written as
\be
\label{n12}
\left|\rho +\s -\frac{r_c H^2}{2} \right|\simeq \m +\frac{H^2}{2\m}  
 \ee
In the case of positive and negative tensions we have
\footnote{To obtain cosmological solutions we must have $\m r_c >1$.}
\ba
\label{n14a}
\s\simeq \m -\rho_0 +\frac{H^2_0}{2\m}(\m r_c +1)\\
\label{n13}
|\s|\simeq \m +\rho_0 -\frac{H^2_0}{2\m}(\m r_c -1 )
.\ea
With expressions (\ref{n14a}) and (\ref{n13})   
 we obtain the Friedmann equations in both models as 
\be
\label{n15}
H^2 =H^2_0 \left[1 + w\left[\left((z+1)^3 -1\right)+ \Omega_r /\Omega_m
\left((z+1)^4 -1\right)\right]\right]
,\ee
where
\begin{eqnarray}
\label{w}
&{}& w=\frac{2}{p^2 (\m r_c +1 )}, \qquad \s>0, \\\nonumber
&{}& w=\frac{2}{p^2 (\m r_c -1)}, \qquad \s<0 .
\end{eqnarray}
For large $z$, such that $ z^4\Omega_r /\Omega_m \gg 1$, condition $H^2
/\m^2 \ll 1$ is satisfied if 
$$
z^4\frac{H^2_0 \Omega_r} {\m^2 p^2\Omega_m }\ll 1
.$$  

Integrating Eq. (\ref{n15}), we obtain the age of the Universe
\footnote{In the cold matter dominated region the the Friedmann equation 
$H^2 =H^2_0\left[1+w\left((z+1)^3 -1\right)\right]$ is integrated
analytically and yields $H_0 t_0
={2}/(3\sqrt{1-w})\ln\left[\sqrt{1-w}+1 )/\sqrt{w}\right]$. 
Numerical results which follow from this formula are very
close to those in the table which include contribution from the radiation
period.}
\begin{center}
\begin{tabular}{|c|c|c|c|c|c|c|c|}\hline
$ w =$ &0.26 & 0.28& 0.3& 0.31 & 0.32 & 0.35& 0.5\\\hline
$t_0 H_0 =$ &1.00 &0.98 & 0.96& 0.96 &0.95  & 0.92& 0.83\\\hline
\end{tabular}
\end{center} 
From Friedmann Eq. (\ref{n15}) we can obtain relation between $w$ and $q_0$.
Taking the time derivative of this equation,
$$
2\dot{H}H=H^2_0 w \dot{z} [3(z+1)^2 + 4(z+1)^3\Omega_r /\Omega_m],
$$ 
and evaluating it at present time $t=t_0$, we obtain
$$
w\simeq\frac{2(1+q_0 )}{3},
$$
where we neglected the small second term.
From the table it is seen that for $w =0.3\pm 0.05$
 the age of the Universe is within the
 observational bounds $t_0 H_0 =0.96 \pm 0.06$.
For $w= 0.3$ the deceleration parameter is $q_0 =-0.55$.

From (\ref{w}) 
 we obtain a relation between parameters of the model
\be
\label{n21}
\frac{\m M^2_{pl}}{M^3}\simeq \frac{2(1+q_0 )}{3\Omega_m}(
\m r_c \pm 1) 
.\ee

\section{Primordial nucleosynthesis}
\subsection{Model with positive tension of the brane}
In this section, comparing predictions of BBN in the standard and
non-standard cosmologies, we obtain constraints on parameters of the
non-standard model. 
First, we study the model with positive
tension of the brane without the $4D$ curvature term in the action.

In the standard cosmology, 
the expressions for the radiation energy density as 
functions of $t(z)$ and of temperature of the Universe $T$ are
\ba
\label{p1}
&{}&\h{\rho} (t) =\h{\rho}_{r0} r (z) z^4 =\frac{\h{\rho}_{r0}/\Omega_r}{4(H_0
t)^2},\\\nonumber
&{}&\h{\rho} (T)=\frac{\pi^2}{30}g_{*}(T) T^4.
\ea
In non-standard model, in the period of late cosmology we have
\ba
\label{p3}
&{}&\t{\h{\rho}}(t)\simeq\frac{\h{\rho}_{r0}}{2(H_0 t)^2}
\frac{p^2 r(z)}{[(1+q_0 )p^2 -3 +4(r(z) -1)\Omega_r/\Omega_m ]},
\\\nonumber
&{}&\t{\h{\rho}} (\t{T})=\frac{\pi^2}{30}g_{*}(\t{T}) \t{T}^4
\ea
where by {\it tilde} we  distinguish 
  the non-standard case.  
The freezing temperature $T_F$
of the reaction $n\leftrightarrow p$ is estimated as a temperature at which
the Hubble parameter $H$ is of order of the reaction rate $G_F^2
T_F^5$ \cite{kolb}. Time dependence of the Hubble 
parameter in both the standard and
non-standard cosmologies is the same $H=1/2t$. 
Using (\ref{p1}) and (\ref{p3}), we obtain the ratio of freezing
temperatures in the standard and non-standard cosmologies
\be   
\label{p4}
\frac{\t{T}_F}{T_F}=\left(\frac{{g}_{*}(\t{T}_F) }{{g}_{*}(T_F )}
\frac{[(1+q_0 )p^2 -3 +4(r(z) -1)\Omega_r/\Omega_m ]}{2p^2 r(z)\Omega_r 
}\right)^{1/6}
.\ee
The mass fraction of  ${}^4 He$ produced in nucleosynthesis of the total
baryon mass is
\be   
\label{p51}
 X_4 = \frac{2(n/p)_f}{(n/p)_f +1} 
,\ee
where the subscript "f" indicates that the ratio is taken at the end of
primordial nucleosynthesis, $(n/p)_f \simeq 1/7$ \cite{kolb}.
 The equilibrium value of the
neutron-proton ratio $(n/p)_{T_F}=\exp{[-(m_n -m_p )/T_F]}$ is very
sensitive
to the value of $T_F$. 
The value of $X_4$ calculated in the standard cosmological model 
\cite{kolb} fits well the cosmological data $X_4 =0.25 \pm 0.01$ 
\cite{part}. 
We constrain parameters of the model requiring that the difference 
between
the calculated values of $X_4$ in the standard and non-standard models
is within the experimental errors.
Under variation of freezing temperature variation of $X_4$ is
\be
\label{p6}
\delta X_4\simeq \frac{2}{((n/p)_f +1)^2} ({n}/{p})_f \ln
({p}/{n})_F\frac{\d T_F}{T_F}.
\ee
Variation of $X_4$ is within the experimental errors for variations of 
freezing temperature in the range
$$
\frac{\d T_F}{T_F}< 0.0255.
$$ 
Comparing $T_F$ and $\t{T}_F$ in (\ref{p4}), we find that variation of 
$X_4$ is within the experimental bounds if
\be
\label{p6a}
\left|\frac{{g}_{*}(\t{T}_F) }{{g}_{*}(T_F )}
\frac{[(1+q_0 )p^2 -3 +4(r(z)-1)\Omega_r/\Omega_m ]}
{2p^2 r(z)\Omega_r }-1 \right| <\e
,\ee
where  $\e=0.164$ 
\footnote{In another context similar estimates were done in 
\cite{bar}.}.
Characteristic temperatures of nucleosynthesis are below the neutrino 
decoupling temperature $T\sim (0.8 \div 1 ) MeV$ and $e^{\pm}$ 
annihilation 
temperature $\sim 0.5 MeV$. At these temperatures the number of 
effective degrees of freedom in the expression of radiation energy 
density is constant $g_* (T)=3.36$ \cite{kolb}. Since in this range the 
number of
effective degrees of freedom does not change, parameter $r$ is also 
constant and equal the present-time value $r=1$.

Substituting in (\ref{p6a}) $p^2\simeq 2M^3/\m M^2_{pl}\Omega_m $ and 
setting $r=1$, we obtain
\be
\frac{4\Omega_r}{3\Omega_m}\left(1-\e \right)
<\frac{\m M^2_{pl}}{M^3} -\frac{2(1+q_0 )}{3\Omega_m}<
\frac{4\Omega_r}{3\Omega_m}\left(1+\e\right).
\ee
which is specification of the relation (\ref{r8}) obtained from the estimate
of the age of the Universe.
 
Now we can verify that in the non-standard model BBN takes place at the
period of late cosmology. 
In the non-standard model, in the radiation-dominated period,
solving  Eq.(\ref{r4}) which interpolates between the
early and late radiation-dominated periods, we find
\be
\label{p8}
z^{-4}\simeq\frac{2[(1+q_0 ){p}^2 -3 ]}{{p}^2} (H_0 t)^2 +
\frac{4 \Omega_{r}}{{p}^2\,\mu \Omega_{m}}H^2_0 t.
\ee
From (\ref{p8}) follows that the transition time $\bar{t}$
 from  $\rho^2$ to $\rho$-dominated law is at $\b{z}\sim (\m^2/H^2_0 
\Omega_{r})^{1/4}$. In the most stringent case, for $\m\sim 10^{-12} 
GeV$, this yields $\b{z}\sim 10^{16}$, which is much larger than 
$z_{BBN}\sim 10^{9\div 10}$. 
From this estimate and (\ref{p8}) it follows  
that the transition time is $\b{t}\sim 1/\m$ 
\footnote{The same estimate follows from condition $\rho (z)\sim \m$.}. 
For the limiting values in the assumed
interval of $\m$ transition times are 
 $\bar{t}\sim 10^{-3} GeV^{-1}$ and $\bar{t}\sim 10^{12} GeV^{-1}$
which  are much
smaller than the  characteristic time of nucleosynthesis
 $1\div 10^2 s$, or
$10^{24\div 26}GeV^{-1}$ estimated in the standard cosmology. 

Let us turn to the model with 4D curvature term included in the action.
In the radiation-dominated period of late cosmology we have
\be
\label{p10}
z^{-4}\simeq\frac{2[p^2 (\m r_c +1 )(1+q_0 )-3 ]}{p^2 (\m r_c +1)} 
(H_0 t)^2 . 
\ee
Following the same steps as in the model with $r_c =0$, we obtain that
production of Helium is within the experimental bounds 
if
\be
\label{p9b}
\left|\frac{p^2 (\m r_c +1 )(1+q_0 )-3 }{2p^2 
(\m r_c +1)\Omega_r} -1\right|< \e
,\ee
where $\e \simeq 0.164$.
From (\ref{p9b}) we find a constraint on $p^2$
\be
\label{p9c}
\frac{6\Omega_r (1+\e)}{(1+q_0 )}<
p^2 (1+\m r_c )(1+q_0 )-3
<\frac{6\Omega_r (1+\e)}{(1+q_0 )},
\ee
and constraint on the scales of the model 
\be
\label{p9a}
\frac{3\Omega_r \Omega_m (1+\e )}{(1+q_0 )^2}<
\frac{M^3 (\m r_c +1)}{\m M^2_{pl}} -\frac{3\Omega_m}{2(1+q_0 )}
< \frac{3\Omega_r \Omega_m (1+\e )}{(1+q_0 )^2}.
\ee
Using the inequalities (\ref{p9c}) we obtain the bounds on $\rho_{w0}$ 
\be
\label{p11}
\frac{H^2_0}{\m^2} \Omega_r (1+\m r_c )
\left(1-\e- \frac{2(1+q_0 )}{3\Omega_m}\right)
<\frac{\rho_{w0}}{\m}<\frac{H^2_0}{\m^2} 
\Omega_r (1+\m r_c )\left(1+\e- \frac{2(1+q_0 )}{3\Omega_m}\right).
\ee
The combination $2(1+q_0 )/3\Omega_m/$ is of order unity, and within the
existing uncertainties of cosmological parameters the sign of $\rho_{w0}$ is
ambiguous. From (\ref{p11}) it follows that  $\rho_{r0}/\m > \rho_{w0}/\m$.

In the model with $r_c \neq 0$ transition from the early to late 
cosmology takes  place at $\bar{z}$ such
that $\rho (\bar{z})< \m$ or 
$$
\bar{z}^4<\frac{\m^2}{H^2_0\Omega_r (\m r_c +1)}
.$$
Requiring that $z_{BBN}\sim 10^{9\div 10}\ll \bar{z}$, we obtain
\be
\label{p16}
\m r_c \ll \frac{1}{\Omega_r 
z_{BBN}^4}\frac{\m^2}{H^2_0}
.\ee
In the most stringent case, for $\m\sim 10^{-12} GeV$,
this is satisfied if
$$
\m r_c \ll 10^{25}.
$$

\subsection{Model with negative tension of the brane}
First, we consider the model without the 4D curvature term in the action.
 Solution
of Friedmann equation in regions (ii) and (iii) of the 
radiation-dominated period is
\be
\label{p13}
z^{-4}\simeq\frac{2(3+(1+q_0 ){p}^2 )}{p^2} (H_0 t)^2 +
\frac{4 \Omega_r}{p^2\,\mu \Omega_m}H^2_0 t.
\ee
 At the characteristic times of nucleosynthesis
 the first term in (\ref{p13}) is dominant, and we can neglect the
second one.

Following the same steps as in the model with positive tension of the brane,
we obtain the ratio of the freezing temperatures  in the
standard and non-standard cosmologies    
\be   
\label{p14a}
\frac{\t{T}_F}{T_F}=\left(\frac{{g}_{*}(\t{T}_F) }{{g}_{*}(T_F )}
\frac{[3+(1+q_0 )p^2]}{2p^2\Omega_r }\right)^{1/6}
.\ee
Because $(3+(1+q_0 )p^2 )/2p^2\Omega_r $ is not a small number, 
the abundance of $He$ calculated in this  model is
 outside the experimental bounds.

In the model with 4D curvature term included in the 5D action we consider
two cases. If $1-\m r_c >0$, the model is similar to the case without 4D
term. As it was discussed above, in this case the model does not describe
nucleosynthesis correctly. If $\m r_c -1 >0$, the model is similar to the
model with positive tension of the brane and can yield correct value for
$X_4$.

\section{Conclusions}
In this paper we discussed a class of warped cosmological solutions in 
5D models with infinite extra dimension.
We considered the models with positive and negative tensions of the
brane. 
We studied separately the  models  with and without 4D
curvature term on the brane included in the 5D action.   
We looked for constraints on parameters of the
models which follow from consistency of predictions of the models with  
the main cosmological data - the
age of the Universe and abundance of ${}^4 He$ produced  in primordial
nucleosynthesis.

For numerical estimates we used the  present-time values of 
the Hubble parameter $H_0$, fractions of cold matter and radiation
$ \Omega_m , \Omega_r$,
deceleration parameter $q_0$ and the mass fraction $X_4$ of ${}^4 He$ of 
 the total barion mass.
The warping scale $\m$ was taken in the interval $10^{-12}-10^3
GeV$.

In the models with positive tension of the brane in both cases with and
without 4D curvature term on the brane the normalized age of the Universe
$t_0 H_0$ was obtained within the observational bound if parameters of the
model are connected by the relation
$$ 
\frac{\m M^2_{pl}}{M^3} \simeq (\m r_c +1)\frac{3\Omega_m}{2(1+q_0 )}.
$$
Comparing abundance of ${}^4 He$ in the standard and non-standard models,
we obtained a further constraint on parameters 
$$
\left|\frac{M^3 (\m r_c +1)}{\m M^2_{pl}} -\frac{3\Omega_m}{2(1+q_0 
)}\right|<C\Omega_r
,$$ 
where $C=O(10)$.
Because $3\Omega_m /2(1+q_0 )\sim 1$, from these relations it follows 
that 
$$
M^3\simeq \frac{\m M^2_{pl}}{\m r_c +1}
.$$

In the case of negative tension of the brane situation is different. With
the input of the current-time cosmological parameters the
model without 4D  curvature term in the action 
does not yield correct age of the
Universe and the abundance of ${}^4 He$. 
In particular, this refers also to the RS-type two-brane models, in 
which the observable brane is that with negative tension \cite{rub}.

The model with 4D curvature term
included in the 5D action can meet the observational data provided
$\m r_c >1$.
The normalized age of the Universe $t_0 H_0$ is obtained of order unity, if
$$
\frac{\m M^2_{pl}}{M^3}\simeq \frac{2(1+q_0 )}{3\Omega_m}(\m r_c -1 ).
$$
From the estimate of the abundance of ${}^4 He$ follows a more precise  
constraint 
$$
\left|\frac{M^3 (\m r_c -1)}{\m M^2_{pl}} - 
\frac{3\Omega_m}{2(1+q_0 )}\right|< C\Omega_r,
$$
where $C\sim 10$.
Further restrictions are obtained, if 
 $M_{pl}$ is the largest scale of the theory, i.e. $M^2_{pl}>r_c M^3$, 
and $3\Omega_m /2(1+q_0 )<1$.
In this case we have
$$
\m r_c <\frac{\m M^2_{pl}}{M^3}< \left(\frac{2(1+q_0 )}{3\Omega_m}
 - C\Omega_r\right)^{-1}(\m r_c -1) ,
$$
or
$$
\m r_c \left(1-\frac{3\Omega_m}{2(1+q_0 )} +C\Omega_r\right)>1
.$$
Because $3\Omega_m /2(1+q_0 )\sim 1$, it follows that in this 
model $\m r_c\gg 1$.

Generally, numerical values of cosmological parameters 
depend on a model in which 
 the experimental data were processed. Recently fits of 
parameters were
performed in one-brane models with positive tension of the brane
\cite{dab,fay,hung1,hung2} 
\footnote {The dark
radiation term was taken as $C/a^{4-\a}$, where $\a=0$ corresponds to the
case of the present paper.}. 

Let us compare the estimates of the present paper in which were used 
parameters of the standard cosmological model  
with those with parameters obtained in these fits.
 Identifying the
corresponding terms in the
Friedmann equation in the form employed in the present paper
$$
H^2 =\s^2 -\m^2 +2\s\rho +\rho^2 +Cz^4
$$
with those in \cite{dab,hung1,hung2} 
$$
H^2 =\frac{\bar{\Lambda}}{3} +\frac{\bar{\kappa}^2\hat{\rho}}{3}
+\frac{\bar{\kappa}^2\hat{\rho}^2}{6\lambda^2} +2m_0 z^4
$$
(because of different normalizations, in
the latter equation we use notations with a bar), we have
$$
\bar{\kappa}^2 =\kappa^2\s,\quad \lambda =\frac{6\s}{\kappa^2},\quad 
\bar{\Lambda}/3 = \s^2 -\m^2,
\quad 2m_0 =C,
$$
Introducing
$$
\bar{\Omega}_\rho =\frac{\bar{\kappa}^2\hat{\rho}_{0}}{3H^2_0},\quad
\bar{\Omega}_\lambda=\frac{\bar{\kappa}^2\hat{\rho}_0^2}{6\lambda H^2_0}, \quad 
\bar{\Omega}_{\bar{\Lambda}}=\frac{\bar{\Lambda}}{3H_0^2 },\quad 
\bar{\Omega}_d
=\frac{2m_0}{H_0^2 },
$$
we obtain the relations
\be
\label{co1}
\bar{\Omega}_\rho =\Omega_m b^2 =\frac{2}{p^2}, \quad 
\bar{\Omega}_{\bar{\Lambda}} = 
\frac{-1 +(1-q_0 )p^2}{2p^2}, \quad 
\bar{\Omega}_d =\frac{C}{H^2_0}=\frac{-3+(1+q_0 )p^2-4\Omega_\rho 
/\Omega_m}{2p^2} 
\ee
Some examples of the best fits of parameters $\bar{\Omega}$ are 
presented in the Table.

\begin{center}
\begin{tabular}{|c|c|c|c|c|c|c|} \hline
ref.  & $\bar{\Omega}_\rho$   & $\bar{\Omega}_{\bar{\Lambda}} 
$&$\bar{\Omega}_d$
&$\bar{\Omega}_\lambda$ & $-q_0$\\
   &     &       &      &       & derived   \\  
\hline
\cite{hung2}& prior(0.15, 0.35)  &      & prior(-0.03,0.07) &     &  \\
  {}   & 0.225  &0.735 & 0.04  &  0    &       0.58  \\
\hline
\cite{fay}&    &    &prior (-0.1, 0.1) &        &    \\
       &0.15&0.80&     0.008      &  0.026   & 0.67\\
 \hline
\cite{fay}&prior (0.2, 0.4)  &      & prior(-0.1, 0.1) &      &     \\
  {}   &          0.29    & 0.78 &  -0.09          & 0.02 & 0.68 \\
\hline 
\cite{dab}&        &        & prior 0 &      &     \\
         & 0.25 & 0.73   &   0     & 0.02 &   0.57\\
\hline
\end{tabular}
\end{center} 

As an example, substituting parameters $\bar{\Omega}_\rho$ and
$\bar{\Omega}_{\bar{\Lambda}}$ from the first line of the table in relations 
(\ref{co1}), we calculate
$p^2 =8.89,\,\, b^2 =0.999$. With these numbers 
 we find $\bar{\Omega}_d =0.04$. 

Let us estimate the BBN constraints on ratio of the dark radiation to 
photon energy density at
$T_F\sim (0.8\div 1\,) MeV$  in the model with $r_c =0$. 
Dark radiation and photon energy density scale with $z$ in
the same way. 
Photon energy density is \cite{kolb}
$$
\rho_\g (z) =\frac{2}{3.36}\rho_{r 0} z^4  . 
$$
Here 
$$
\rho_{r 0} =\frac{\Omega_r}{\Omega_m}\frac{H^2_0}{p^2\m}.
$$
The bound  (\ref{p11}) can be rewritten as
$$
\frac{H^2_0}{p^2\m^2}\frac{3\Omega_r }
{1+q_0}\left(1-\e -\frac{2(1+q_0 )}{3\Omega_m}\right)
<\frac{\rho_w}{\m}<\frac{H^2_0}{p^2\m^2}\frac{3\Omega_r }
{1+q_0}\left(1+\e -\frac{2(1+q_0 )}{3\Omega_m}\right) 
.$$
For the  ratio of energy densities we obtain 
$$
3.36\frac{3  \Omega_m   }{2(1+q_0 ) }
\left(1-\e -\frac{2(1+q_0 )}{3\Omega_m}\right)<
\frac{\rho_w}{\rho_\g} <3.36\frac{3 \Omega_m   }{2(1+q_0 ) }
\left(1+\e -\frac{2(1+q_0 )}{3\Omega_m}\right)
.$$
  With $\e=0.164$ and 
$2(1+q_0)/3\Omega_m =1.00$ we have 
$$
\left|\frac{\rho_w}{\rho_\g}\right|<0.55
,$$
to compare with the bound $-0.4<\rho_w/\rho_\g < 0.1$  of 
\cite{ichiki} . 

Using parameters of best fit of \cite{hung2} (the first
line of the table) we obtain $2(1+q_0)/3\Omega_m =1.24$. In this case
$  -1.1 <\rho_w /\rho_\g < -0.2$ to compare with the result of 
\cite{hung2}
$-1.32<\rho_w /\rho_\g < 0.34$.

\end{document}